\documentclass[%
 reprint,
superscriptaddress,
longbibliography,
nofootinbib,
 amsmath,amssymb,
 aps,prl,
]{revtex4-2}

\usepackage{graphicx}% Include figure files
\usepackage{dcolumn}% Align table columns on decimal point
\usepackage{bm}% bold math
\usepackage{hyperref}% add hypertext capabilities
\newcommand\bw{\begin{widetext}}
\newcommand\ew{\end{widetext}}

\newcommand{\dd}{\,\partial}
\newcommand{\bb}{\,\square}

 \def\be{\begin{equation}}
\def\ee{\end{equation}}
 \def\ba{\begin{align}}
\def\ea{\end{align}}
\def\bea{\begin{eqnarray}}
\def\eea{\end{eqnarray}}

\begin{document}

%\preprint{APS/123-QED}

\title{Anomaly-free scale symmetry and gravity}% Force line breaks with \\

\author{Mikhail Shaposhnikov}
\email{mikhail.shaposhnikov@epfl.ch}
 \affiliation{Institute of Physics, 
\'Ecole Polytechnique F\'ed\'erale de Lausanne (EPFL), CH-1015 Lausanne, Switzerland}%Lines break automatically or can be forced with \\
\author{Anna Tokareva}%
 \email{a.tokareva@imperial.ac.uk}
 \affiliation{current affiliation: Theoretical Physics, Blackett Laboratory, Imperial College London, SW7 2AZ London, U.K.}
\affiliation{Department of Physics, University of Jyv\"askyl\"a, P.O.Box 35 (YFL), FIN-40014, Finland }
\affiliation{Helsinki Institute of Physics (HIP), University of Helsinki, P.O. Box 64, 00014, Finland}

\date{\today}% It is always \today, today,
 % but any date may be explicitly specified

\begin{abstract}
In this Letter, we address the question of whether the conformal invariance can be considered as a global symmetry of a theory of fundamental interactions. To describe Nature, this theory must contain a mechanism of spontaneous breaking of the scale symmetry. Besides that, the fundamental theory must include gravity, whereas all known extensions of the conformal invariance to the curved space-time suffer from the Weyl anomaly. We show that conformal symmetry can be made free from quantum anomaly only in flat space. The presence of gravity would reduce the global symmetry group of the fundamental theory to the scale invariance only. We discuss how the effective Lagrangian respecting the scale symmetry can be used for the description of particle phenomenology and cosmology.
\end{abstract}

%\keywords{Suggested keywords}%Use showkeys class option if keyword display desired
\maketitle

%\tableofcontents

{\em Introduction.}---What is the global symmetry of Nature? In the {\em absence of gravity} the most obvious answer to this question is given by special relativity and is associated with the Poincar\'e group. The Poincar\'e transformations -- time and space translations, together with Lorentz boosts and rotations -- do not change the form of Maxwell equations. As was noted a long time ago, the Maxwell equations without external currents have a wider symmetry group \cite{Bateman:1909,Cunningham:1910} - the 15 parameters conformal invariance, containing in addition to ten Poincar\'e generators, four special conformal transformations, and dilatations. Dilatations change the length of the rulers, while special conformal transformations can bend the lines but do not alter the angles between them. Could it be that the symmetry of all interactions is conformal?

The answer is certainly ``no'' if the ground state of the theory respects the conformal invariance. The theories enjoying the conformal symmetry -- CFTs -- do not contain any intrinsic mass scale and do not have well-defined particle states, in contrast with observations. However, if the conformal invariance (CI) is spontaneously broken, $\langle O\rangle\neq 0$ (here $O$ is some operator with non-zero mass dimension), the scale appears, massive particle excitations show up, and the resulting theory may happen to be relevant for the description of all particle interactions. 

The theories with exact but spontaneously broken conformal invariance are very interesting from many points of view:\\
- If the mass of the Higgs boson in the Lagrangian of the Standard Model is put to zero, this theory becomes conformally invariant at the classical level. Perhaps, this is a key for an explanation of why the Fermi scale is much smaller than the Planck scale \cite{Wetterich:1983bi,Bardeen:1995kv,Shaposhnikov:2018xkv,Shaposhnikov:2018jag,Shaposhnikov:2020geh}.\\
- The spontaneous breaking of CI leads to the existence of massless Goldstone particle - dilaton. The degeneracy of the vacuum ensures that the energy of the ground state is equal to zero \cite{Amit:1984ri,Einhorn:1985wp,Rabinovici:1987tf,Shaposhnikov:2008xi} - an intriguing fact given the astonishing small value of the observed cosmological constant (dark energy). \\

- The Poincar\'e group has more representations for the massless states than were observed in Nature. In particular, particles with continuous spin can exist (a textbook discussion can be found in \cite{Weinberg:1995mt}, see also \cite{Schuster:2014hca,Buchbinder:2018yoo}) if the Nature is just Poincar\'e invariant. The puzzling absence of these states gets explained as the conformal symmetry does not allow for this kind of irreducible representations \cite{Mack:1969dg, Mack:1969rr}. 

 All the points above (including those about the structure of the ground state) except the last one, are also true if the conformal invariance is replaced by a weaker requirement of the scale invariance.

The list of known UV complete CFTs is very limited and contains only supersymmetric theories such as N=4 Yang-Mills or fishnet CFT \cite{Karananas:2019fox}. The relevance of these theories for the description of Nature is obscure. Coming from another end -- low energies, it is possible to construct phenomenologically viable (e.g. containing just the Standard Model) quantum {\em effective field theories} with exact but spontaneously broken conformal symmetry \cite{Wetterich:1987fm, Wetterich:1987fk},\cite{Shaposhnikov:2008xb,Shaposhnikov:2008xi,Gretsch:2013ooa,Garcia-Bellido:2011kqb,Bezrukov:2012hx} (for a review see \cite{Wetterich:2019qzx}).  These theories are non-renormalizable but weakly coupled below the energy scale of CI breaking. One may hope that their UV limit is given by some hypothetical well-defined CFT.

Suppose that indeed such a theory can be constructed in Minkowski space-time. Will it survive if gravity is added in such a way that the flat space remains a solution to the Einstein equations? In more formal terms, the question can be formulated as follows. The spectrum of spontaneously broken conformally invariant Minkowski theory contains several massive particles and one massless particle - the dilaton\footnote{The realistic theory would also contain a massless photon. The addition of the massless vector field can be made without difficulties and thus is not considered in what follows.}. In the low energy domain, all massive particles can be integrated out, leading to the most general CI action for the dilaton. Consider now an arbitrary action containing two massless particles - a scalar field and graviton, invariant under local coordinate transformations (Diffs).  Actions with this field content can have additional symmetries restricting their form. Can one find %the additional symmetry such that 
such symmetry transformations that if the general metric\footnote{While the dilatation symmetry can be easily defined in a theory with gravity, this is not so for the special conformal transformations.} is replaced by the non-dynamical Minkowski one, the resulting scalar action will be the most general conformally invariant one? Note that this problem is different from that of the construction of anomaly-free Weyl invariant theory of gravity, addressed in \cite{Fradkin:1983tg}.

It is often stated (see, e.g. \cite{Luty:2012ww}) that a natural extension of the conformal symmetry of flat space-time to curved space-time is the Weyl symmetry, constituting in the replacement $g_{\mu\nu}\rightarrow \Omega^2 g_{\mu\nu}$, where $ \Omega$ is an arbitrary function of space-time coordinates. Physically, Weyl invariance is the local freedom of changing the length units. Now, the Weyl symmetry conflicts with quantum theory because of the Weyl anomaly: the classical theory invariant with respect to Weyl transformations loses this property when quantum effects are accounted for (see a review \cite{Duff:1993wm} and references therein) if the requirement of the locality of the effective action is imposed. Even the approach suggested in \cite{Englert:1976ep}  and developed in \cite{Shaposhnikov:2008xi} based on defining the theory as Weyl-invariant in $D=4-2\epsilon$ dimensions by identifying the mass parameter of dimensional regularisation $\mu$ with the dilaton field $\phi$ does not help to avoid the Weyl anomaly with $D$ even (and in particular, in $D=4$) \footnote{However, within this construction the scale \cite{Shaposhnikov:2008xb} and conformal \cite{Gretsch:2013ooa, Shaposhnikov:2022zhj} symmetries can be kept at the quantum level in the {\em flat space}.}.

This Letter aims to demonstrate that the only finite subgroup of the Weyl group, which can be made anomaly-free in any curved spacetime corresponds to the dilatations. The conformal symmetry which was held in the flat space at the quantum level is lost, once gravity is taken into account. The Weyl anomaly in the generic spacetime cannot be cancelled by the proper choice of counterterms and conformal symmetry reduces to the scale symmetry. The latter can be kept anomaly-free at the quantum level.

{\em General structure of the dilaton effective action in Minkowski space-time.}---Though our interest is in the theory in 4-dimensional space-time it would be convenient to consider arbitrary $D$ dimensions. The dilaton field is denoted by $\phi$ and is assumed to have a canonical kinetic term $\partial_\mu\phi \partial^\mu\phi$. The mass dimension of $\phi$ is $\Delta=(D/2-1)$  and equal to $1$ for $D=4$, allowing for spontaneous breaking of conformal symmetry by the dilaton vacuum expectation value.  The most general local conformally invariant quantum effective action which includes all quantum fluctuations can be constructed with the use of the following two ingredients \cite{Shaposhnikov:2022zhj}. The first one is the combination 
\begin{equation}
\label{boxes}
O_N=\phi^{\frac{2(N+1)}{2-D}}\square^N \left(\phi^{\frac{D-2N}{D-2}}\right)~,
\end{equation}
which changes like $\phi$ under special conformal transformation, $\phi(x)\to \phi'(x')=\Omega^{-\Delta}\phi(x)$, where $\Omega=(1+2a_{\mu}x^{\mu}+a^2 x^2)^{-1}$ and $a_\mu$ is an arbitrary $D$-vector. The coordinates are to be transformed as
\begin{equation}
x_{\mu}\rightarrow x'_{\mu}=\frac{x_{\mu}+a_{\mu}x^2}{1+2 a_{\mu}x_{\mu}+a^2 x^2}.
\end{equation}
Here $\square$ is the D-dimensional Laplacian and $N$ is an integer number. The second ingredient is the differential operator 
\begin{equation}
\label{oper}
\hat{O}_N=\phi^{\frac{2(N+1)}{2-D}}\bb^N ~,
\end{equation}
which can be applied to any scalar operator with mass dimension $D/2-N$.

The conformally invariant Lagrangian can be written as a sum of all combinations of the form
\begin{equation}
\label{conf_O}
\phi^q\hat{O}_{N_1}[\phi^{\alpha_1} O_{m_1}...O_{m_k}]...\hat{O}_{N_p}[\phi^{\alpha_p} O_{s_1}...O_{s_l}]~,
\end{equation} 
where the powers $\alpha_i$ are fixed in such a way that the mass dimension of the operators in square brackets is equal to $D/2-N_i$ and the power $q$ is singled out from the requirement that the action is dimensionless. Not all of these operators are independent, some of them can be related to each other via integration by parts. The key observation in the procedure of construction of conformal operators is that the power of the $\square$ operator has to be properly adjusted to the conformal weight of the operator which is differentiated. To the best of our knowledge expression (\ref{conf_O}) is new and has not appeared in the literature.

The conformally invariant operators can be classified by the total number of derivatives (in what follows we take $D=4$). There is one operator $\phi^4$ without derivatives, one operator $\phi\bb\phi$ (the kinetic term) with two derivatives, two operators with 4 derivatives
(there is one more operator in scale-invariant but not CI theory, $O_3=\frac{1}{\phi^4}(\dd_{\mu}\phi)^4$), 
\begin{equation}
\label{4der}
Q_1=\tau\bb^2\tau,~~Q_2=\frac{1}{\phi^2}(\bb\phi)^2~,
\end{equation}
(here $\tau= \log(\phi/\mu)$ \footnote{Note that the action does not depend on the dimensionful parameter $\mu$, as it disappears after integration by parts.}),  4 operators with 6 derivatives (while there are 7 scale-invariant ones)\footnote{According to \cite{Luty:2012ww}, for unitary renormalisable perturbative theories in flat space-time the scale invariance implies conformal symmetry. It may be not so if the requirements of renormalisability and perturbativity are removed, leading to the mismatch between the number of scale-invariant and conformally invariant higher-dimensional operators.}, etc. Note that the operator $Q_1$ (it will play an important role below) can be derived as the formal limit $Q_1=\lim_{D\to4}\left(\frac{2}{D-4}\bb\phi^{\frac{D-4}{2}}\right)^2$. 

{\em Weyl anomaly versus Weyl symmetry.}---There is yet another way to construct the local dilaton effective action which is based on the following observation. Take an arbitrary Diff-invariant action constructed from the metric only (i.e. consider pure gravity) and replace the metric $g_{\mu\nu}$ by $g_{\mu\nu}\phi^2/M_P^2$, where $M_P$ is any scale normally taken to coincide with the Planck mass. One gets in this way a scalar-tensor gravity which is Weyl-invariant (under transformation $g_{\mu\nu}\rightarrow \Omega^2 g_{\mu\nu},~\phi\to\Omega^{-1}\phi$) by construction. Now, it is obvious that a Weyl-invariant theory is automatically conformal invariant if the metric is taken to be flat because the metric rescaling can be substituted by the suitable transformation of coordinates of the flat space. This method has been used for the construction of the dilaton action in spaces with different dimensions (see, e.g. \cite{Komargodski:2011vj,Luty:2012ww,Baume:2013ika,Osborn:2015rna}). 

This procedure leads to the impression that the natural extension of the conformal invariance to curved space-time is the Weyl symmetry. However, the Weyl symmetry happened to be anomalous \cite{Duff:1993wm}. The reason why this is the case is based on a simple counting of the available Diff-invariant operators. For $D=4$, and for four derivatives, these are: $R^2$, $R_{\mu\nu}R^{\mu\nu}$, $W^2=W_{\mu\nu\rho\sigma}W^{\mu\nu\rho\sigma}$, and $\bb R$, where $R$ is the scalar curvature, $R_{\mu\nu}$ and $W_{\mu\nu\rho\sigma}$ are the Ricci and Weyl tensors respectively. Out of these four operators, $\bb R$ is a full derivative and thus it cannot be used for construction of the conformal action in flat space, $W^2$ is Weyl-invariant and thus it does not lead to any non-trivial scalar action, whereas the combination $E_4=R_{\mu\nu\rho\sigma}R^{\mu\nu\rho\sigma}-4R_{\mu\nu}R^{\mu\nu}+R^2$ is the Euler density which is a surface term\footnote{See \cite{Yale:2011usf} for discussion how the invariant $\sqrt{-g}E_4$ can be represented as a total derivative in 4 dimensions.}.

In $D=4$ only one conformally invariant operator, containing four derivatives - $Q_1$ - failed to be constructed by the above procedure. However, as it follows from the a-theorem \cite{Komargodski:2011vj} this operator must be present in the effective action of the dilaton obtained in any unitary theory. All CI operators containing 2 derivatives, as well as 6 and more derivatives can be found in this way. A similar story happens in higher even dimensions $D=2k,~k\geq 2$, namely the CI operators $\tau\bb^k\tau$ cannot be derived from local Diff-invariant pure gravity action, because of the existence of topological Euler densities in $2k$-dimensional space. (However, {\em non-local} operators providing the Weyl invariant action can be constructed \cite{Fradkin:1983tg, Antoniadis:1991fa}.) This makes it clear that the extension of the flat space CI to the curved space cannot be the Weyl symmetry if the dimension of space-time is even and the locality of the action is imposed.

{\em How to extend the conformal symmetry to the curved space?}---The conformal group in flat space depends on a finite number of parameters, whereas the Weyl symmetry is local and thus is controlled by an arbitrary function of space. This poses the question of whether one can find a finite subgroup of the Weyl group which is anomaly-free and matches the flat-space conformal symmetry. 

To analyse this problem we note that the correspondence between the number of conformal operators with 4 derivatives in flat space and the number of Diff-invariant operators in curved space is restored if $D \neq 4$. It is customary to take formally $D\neq 4$, as in dimensional regularisation, and consider eventually the limit $D\to 4$. Perhaps, the most compact Weyl-invariant extension of the $Q_1$-type flat space dilaton action to curved space is
\begin{widetext}
\begin{equation}
\label{div}
\int d^4 x \left(\tau\bb^2\tau\right) \to S=\lim_{D\to 4} \int d^D x \sqrt{-g}\left[\tau\Delta_4\tau+
2\tau \left(-\frac{1}{6} \bb R +\frac{1}{4} E_4\right)+\frac{R^2}{36} + L_{anom}\right]~, 
\end{equation}
\end{widetext}
where $\Delta_4$ is the so-called Riegert operator \cite{Fradkin:1981jc,paneitz,Riegert:1984kt} which is conformally invariant for a scalar field with mass dimension zero, 
\[
 \Delta_4=\bb^2+2R^{\mu\nu}\nabla_\mu\nabla_\nu-\frac{2}{3}R\bb+\frac{1}{3}(\nabla^\mu R) \nabla_\mu
\]
and
\begin{equation}
\label{anom}
 L_{anom}=\frac{E_4}{2(D-4)}~.
\end{equation}
The action $S$ can be modified by adding the surface or Weyl-invariant terms such as $\bb R$ or $W^2$. The appearance of the formally singular at $D\to 4$ term in (\ref{div}) is exactly the manifestation of the Weyl anomaly \footnote{Notice that for the field with zero mass dimension the singularity in $D=4$ appears starting from Weyl-invariant extension of $\bb^3$ operator \cite{Karananas:2015ioa}}. There is no way to construct a local operator in $D=4$ with Weyl transformation properties of $ L_{anom}$.

The infinitesimal Weyl transformation $\Omega = 1+\omega,~\omega \ll 1$ of the anomaly term is $\delta_\omega L_{anom}=\frac{1}{2}E_4\omega$. It is finite and non-zero when $D\to 4$. The anomaly-free subgroup is singled out by the requirement that this variation for $\omega$ belonging to this subgroup must be zero for an arbitrary metric.  For this to happen  $E_\omega= \int d^4 x \sqrt{-g}E_4\omega$ must be a surface integral.  Our main observation is that this is the case  only if 
\begin{equation}
\label{SWS}
 \nabla_{\alpha}\nabla_{\beta} \,\omega=0~, 
\end{equation}
where $\nabla_{\alpha}$ is a covariant derivative. To see this, we consider the variation of $E_\omega$ with respect to the metric, $g_{\mu\nu}\to g_{\mu\nu} +h_{\mu\nu}$,
\begin{equation}
    \delta E_\omega = \int d^4 x \sqrt{-g}h_{\mu\nu}\Sigma^{\mu\nu\alpha\beta}\nabla_{\alpha}\nabla_{\beta}\,\omega.
\label{int}
\end{equation}
A tedious but straightforward computation gives
\begin{widetext}
\begin{equation}
\Sigma^{\mu\nu\alpha\beta}=2R(g^{\alpha\mu}g^{\beta\nu}-g^{\alpha\beta}g^{\mu\nu})+4 R^{\mu\nu}g^{\alpha\beta}+4g^{\mu\nu}R^{\alpha\beta}-8 g^{\mu\beta}R^{\alpha\nu}-4 R^{\mu\alpha\nu\beta}~.
\end{equation}
\end{widetext}
 In a general metric and for arbitrary $h_{\mu\nu}$, the integral (\ref{int}) is zero only if eq. (\ref{SWS}) is satisfied.

The equation \eqref{SWS} can easily be solved in the flat space and the solutions would correspond to the familiar conformal transformations, $\omega =c(1+2a_{\mu}x^{\mu})$. This way, one can see that conformal symmetry can be made anomaly-free in a flat space.

Are there any solutions to the equation \eqref{SWS} on top of the generic gravitational background? Taking the covariant derivative of \eqref{SWS} one obtains a commutator
$[\nabla_{\gamma}\nabla_{\beta}]\nabla_{\alpha}\omega=R_{\beta\gamma\lambda\alpha}\nabla^{\lambda}\omega=0$. Thus, all solutions of \eqref{SWS} must satisfy $R_{\alpha\beta}\nabla^{\beta}\omega=0$. If the metric is not flat this equation has only the trivial constant solution which corresponds to the dilatations. Thus, we see that, among all possible Weyl transformations, only the scale symmetry can be kept non-anomalous in the presence of gravity. This conclusion holds even if one uses the renormalization procedure which preserves the scale symmetry. In the flat space, the anomaly-free symmetry group is larger and corresponds to the full conformal symmetry, however, the special conformal transformations cannot be matched to some subgroup of Weyl transformations.

A remark is now in order. There are two other finite subgroups of the Weyl symmetry which were considered in the literature \cite{Iorio:1996ad,Karananas:2015ioa,Edery:2014nha,Edery:2015wha}. A covariant extension of the conformal transformations in curved space can be defined \cite{Iorio:1996ad,Karananas:2015ioa} (see also \cite{Rychkov:2016iqz}) via the Killing vector $\xi_{\mu}$ satisfying the equation
\begin{equation}
\label{killing}
 \nabla_{\mu}\xi_{\nu}+\nabla_{\nu}\xi_{\mu}=\frac{2}{D}g_{\mu\nu}\nabla^{\alpha}\xi_{\alpha}~.
\end{equation}
The corresponding infinitesimal Weyl factor is given by $\omega=\nabla_{\alpha}\xi^{\alpha}$. One can check that for this choice $\nabla_{\alpha}\nabla_{\beta}\,\omega \neq 0$ for generic non-flat metric \cite{Shaposhnikov:2022zhj}, meaning that this symmetry is anomalous.

Another related possibility for the extension of the conformal symmetry named restricted Weyl transformations was studied in \cite{Edery:2014nha,Edery:2015wha}. In these works, it was shown that in $D=4$ the transformations satisfying $\Box \Omega=0$ form a subgroup in the group of Weyl transformations. This symmetry is also anomalous, only a subgroup of it with $\Omega=$const can survive at the quantum level.

{\em Scale-invariant action.}---Here we present the first few operators in the derivative expansion of the Lagrangian for the dilaton and gravity in $D=4$ scale symmetry. All the terms that could be written at the level of zero and two derivatives are
\begin{equation}
 \label{SWeyl-inv2}
 S=\int{d^4 x\sqrt{-g}\left[-\frac{1}{2}\zeta\phi^2 R+\frac{1}{2}\kappa(\partial_{\mu}\phi)^2-\frac{\lambda}{4}\phi^4\right]},
\end{equation}
where $\zeta$ and $\lambda$ are arbitrary constants  and $\kappa=\pm 1$ (note that Weyl invariance would impose a specific value for the non-minimal coupling $\zeta=-1/6\kappa$).\footnote{ At first sight, the value $\kappa=-1$ corresponds to the presence of the ghost particle in the theory. However, if $0 < \zeta < 1/6$, the theory is ghost-free both in the scalar and gravitational sectors, as can be seen in the Einstein frame.} Making a transition to the Einstein frame one can see that the combination $\frac{\lambda M_P^4}{\zeta^2}$ is the vacuum energy. In these terms, the cosmological constant problem is converted to the question of why $\lambda/\zeta^2\ll 1$.

 The general Lagrangian invariant under the scale transformations and respecting parity at the level of four derivatives can be written as
\begin{widetext}
\begin{equation}
\label{SWeyl-inv4}
\begin{split}
 S=\int d^4 x \sqrt{-g} \left[A R\Box \tau+B R(\partial_{\mu}\tau)^2)+C R^{\mu\nu}\partial_{\mu}\tau\partial_{\nu}\tau + E\tau E_4 +F R^2+G W_{\mu\nu\lambda\rho}^2+H((\partial_{\mu}\tau)^2)^2\right.+\\
 \left.+J(\Box \tau)^2+K (\Box \tau+(\partial_{\mu}\tau)^2)^2\right]~ .
 \end{split}
\end{equation}
\end{widetext}
Here $A,B,C,E,F,G,H,J$ and $K$ are arbitrary constants, $G_{\mu\nu}$ is the Einstein tensor, $\tau=\log({\phi/\mu})$ with $\mu$ being an arbitrary scale \footnote{Nothing depends on this scale in perturbative computations as it disappears after integrating by parts. This may not be the case for non-perturbative effects associated with configurations with non-zero Euler characteristics coming from the term containing $E$.}. The field $\tau$ transforms under the dilatations as $\tau\rightarrow \tau-\omega$. 

The second line of \eqref{SWeyl-inv4} contains only the operators which are allowed by the conformal symmetry in the flat space limit. Given the fact that the conformal symmetry is broken by gravity, the operators in the first line are expected to be suppressed by the Planck scale. This is not necessarily the case for the two conformal operators since they have an enhanced symmetry in the flat space limit.

The structure of the action given by (\ref{SWeyl-inv2},\ref{SWeyl-inv4}) allows also us to clarify the situation with the energy-momentum tensor. It is well known \cite{Callan:1970ze} that in CFT in flat space it is possible to define an (improved) stress-energy tensor $T_{\mu\nu}$ with zero trace, $T_\mu^\mu=0$. If the theory were Weyl invariant, this relation would remain in force in the gravitational background. However, when the quantum corrections are incorporated in a Weyl-invariant way in $D$-dimensions as in \cite{Englert:1976ep,Shaposhnikov:2008xi}, and the limit $D\to 4$ is taken, $T_\mu^\mu$ receives several contributions containing Diff invariants $W^2,~E_4,~R^2$ and $\bb R$ \cite{Duff:1993wm}. One of them ($E_4$, the so-called a-anomaly) cannot be removed by the Weyl-invariant counter-term, signalling that the Weyl symmetry is anomalous \footnote{Note that the term $W^2$ representing the so-called c-anomaly can be taken away by the Weyl-invariant counter-term  $\phi^{\epsilon} W^2/ \epsilon$ \cite{Englert:1976ep}.}. The scale symmetry in curved space {\em does not impose} that the trace of stress-energy tensor is zero, only a weaker condition $\int d^4x \sqrt{-g} T_\mu^\mu \omega=0$, where $\omega=$const  must be satisfied.

Talking about phenomenology and cosmology, the graviton-dilaton action \eqref{SWeyl-inv2} can be complemented by all the fields of the Standard Model or $\nu$MSM \cite{Asaka:2005an,Asaka:2005pn} in a scale-invariant way, the explicit equations can be found in \cite{Shaposhnikov:2008xb}. Our findings reveal that scale invariance implies conformal invariance at the level of the lowest order action in the flat space, however, when gravity is included, the symmetry of the Higgs-dilaton action reduces back only to the subgroup of dilatations. We expect that the higher-order operators with conformal symmetry in the flat space are less suppressed than those with only scale symmetry. This hierarchy is a key observation that is relevant for phenomenology since the cutoff scale in the scalar sector in the Higgs-dilaton model is known to be much less than the Planck scale \cite{Garcia-Bellido:2011kqb}.

It has been shown that this Higgs-dilaton Lagrangian can solve all the observational problems of the Standard Model (such as inflation, neutrino masses, baryon asymmetry of the Universe, and Dark Matter), provided the scale symmetry is spontaneously broken (for discussion of inflation see \cite{Garcia-Bellido:2011kqb}, and for review of other problems \cite{Boyarsky:2009ix}). It is of crucial importance that the non-minimal couplings of the Higgs and dilaton fields to the Ricci scalar are not constrained by any value ($-\frac{1}{6}$ for the Weyl symmetry). We note also that the massless dilaton does not lead to the fifth force and thus is harmless from the experimental point of view \cite{Wetterich:1987fm, Wetterich:1987fk,Shaposhnikov:2008xb,Ferreira:2016kxi}. 

{\em Conclusions.}---In this Letter, we found that the anomaly-free extension of the conformal symmetry of flat space-time to curved space-time reduces to the scale symmetry. If this symmetry is spontaneously broken, it is consistent with all available experiments and observations and thus may play a role as the global symmetry of Nature.  All other subgroups of Weyl transformations cannot be made non-anomalous even if the renormalization procedure is preserving the Weyl symmetry.
Thus, in this framework, the conformal symmetry appears to be an accidental symmetry of the Standard Model with the dilaton, being exact only for the action containing at most two derivatives.

Several important and very challenging problems need to be solved to complete the effective field theory picture discussed in this Letter. One of the problems is connected to the tiny value of the cosmological constant which requires an extremely small value of the dilaton self-coupling $\lambda$. Yet another difficult task is building the connection between the low-energy effective theory with spontaneously broken scale symmetry with some theory at high energy where the scale invariance is restored. 

\begin{acknowledgments}
We thank Georgios Karananas, Alexander Monin and Arkady Tseytlin for very useful discussions and comments on the manuscript. We are indebted to Andrei Mikhailov for pointing out an error in the first version of this paper, leading to the revision of its part. A.T. thanks EPFL, where a part of this work was done, for hospitality. This work was supported by ERC-AdG-2015 grant 694896 and by the Swiss National Science Foundation Excellence grant 200020B 182864. The work was also supported by the Academy of Finland grant 318319. 
\end{acknowledgments}

\end{document}